\documentclass[iop]{emulateapj} % read by: ACROREAD; emulateapj-rtx4
\shortauthors{Sekanina \& Kr\'olikowska}
\shorttitle{Sungrazing Comet C/2026 A1 (MAPS)}
\slugcomment{Version \today }

\begin{document}
\title{Orbital Motion and History of Sungrazing Comet C/2026 A1 (MAPS)}
\author{Zdenek Sekanina$^1$ and Ma{\l}gorzata Kr\'olikowska$^2$}
\affil{$^1$La Canada Flintridge, California 91011, U.S.A.; {\sl
 ZdenSek@gmail.com}\\$^2$Centrum Bada\'n Kosmicznych Polskiej Akademii
 Nauk, 00-716 Warszawa, Poland; {\sl mkr@cbk.waw.pl}}

\begin{abstract} % maximum length = 1920 characters
We report results of our extensive computations of the orbital motion of
comet C/2026~A1, including~its integration back to the fourth century and
parallel integration of 1000 virtual clones.~We focus~on~this Kreutz
sungrazer's apparent association with the daylight comets in AD~363,
recorded by Ammianus Marcellinus.  We show that the derived time of the
previous perihelion is strongly affected by erratic outgassing-driven
nongravitational forces and depends on the chosen set of observations.
We find that the previous perihelion was reached within about 1$\sigma$,
or some $\pm$15~years, of AD~363 only when we use the~observations up to
2026 February~9--12, the time of a major anomalous feature~on~the~comet's~light
curve.  We suggest that three bright dwarf sungrazers detected between
2026 March~31 and April~12 were almost certainly distant companions,
offering evidence that C/2026~A1 was part of a larger~object, which
separated at perihelion in AD~363 from what appears to have been the
ancestor of the~Great March~Comet of 1843 and then fragmented far from
the Sun after aphelion.  This scenario~is~supported by required orbital
similarity and suggests that comet C/2026~A1 was member of Population~I.
\end{abstract}
\keywords{individual comets:\ 372\,BC, 363, X/1106\,C1, C/1843\,D1, C/1882\,R1, C/1963\,R1, C/1965\,S1, C/1970\,K1, C/2011\,W3, C/2026\,A1 and companions; methods:\ data analysis\vspace{-0.1cm}}
\section{Introduction} % Section 1
\vspace{-0.04cm}
Shortly after its discovery on 2026 January 13 (Maury 2026), comet
C/2026~A1 was identified as a~\mbox{member}~of the Kreutz system of
sungrazing comets (Green 2026a) and positions from nearly two dozen
pre-discovery CCD images taken at three stations of the {\it Asteriod
Terrestrial-Impact Last Alert System\/} (ATLAS) search project
between 2025~December~18 and 2026~January~10 were reported by Deen
(2026).  The last exposure\vspace{-0.015cm} was taken from the ground
on 2026~March~28, a week before perihelion,\footnote{See {\tt
% so that the object was
% under observation over a period of 100~days.\footnote{See {\tt
https:/$\!$/minorplanetcenter.net/db\_search}.} so that the object
was under observation over a period of 100~days.  With approximately
1000~images altogether taken and measured at nearly 80 observatories
all over the world, it was expected that a high-quality orbit with
an accurate enough orbital period should be determined to ascertain
the sungrazer's history and its place in the overall architecture
of the Kreutz system.

As described in some detail in this paper, the initial optimistic
outlook has recently been revised for a variety of reasons.  One
of them is the size of the nucleus.  The reported apparent
magnitudes of the comet, 20.4 at a heliocentric distance of 2.46~au
and a geocentric distance of 1.77~au (at the time of the first
pre-discovery observation) and $\sim$18 at, respectively, 2.06~au
and 1.44~au (at discovery), already suggested the comet's modest
activity at best.  This skepticism was confirmed, when the diameter
of the nucleus was estimated at 0.4~km by Zhang et al.\ (2026),
based on their infrared imaging of the comet with the {\it James
Webb Space Telescope\/} (JWST) on 2026 February~6--7.  A nucleus of
these dimensions is subjected to an outgassing-driven nongravitational
acceleration, whose magnitude is sufficiently high to measurably
affect the comet's orbital motion, introducing uncertainties into
the computations, and downgrading the accuracy of an orbital-period
determination.

Another major complication has arisen from the detection of dwarf
Kreutz sungrazers that in all probability accompanied the comet on its
journey to perihelion.~Impact of the fragmentation episode(s) on the
comet's motion in general and the orbital period in particular can
be expected.  At least three~fairly bright presumed companions have
been located in numerous images taken on 2026 March~31, April~8,
and April~12 with the C2 coronagraph on board the {\it Solar and
Heliospheric Observatory\/}\,(SOHO)\,and the \mbox{CCOR-1} \mbox{coronagraph}
on~board~the {\it Geostationary Operational~Environmental Satellite--19\/}
(GOES--19).  The companions undoubtedly disintegrated before
perihelion, estimated to have taken place, respectively, 4.3~days
before, 4.4~days after, and 7.8~days after the comet's perihelion time.
While one cannot prove it without careful analysis of their motions,
the genetic relationship of these objects with the comet is virtually
certain, because an arrival rate of one fortuitous bright dwarf SOHO
sungrazer per four days has been unheard of.\footnote{R.\ Kracht
({\tt http:/$\!$/www.rkracht.de/soho/bright/bright.htm}) and
independently Knight et al.\ (2010) estimated an average rate of
SOHO Kreutz sungrazers brighter at maximum than apparent magnitude~3
to amount to about 0.8 per year (or one per 460~days; a rate lower by
a factor of $>$100).  Sekanina \& Kracht~(2013) investigated a ``swarm''
of such bright dwarf sungrazers in late~2010, whose peak rate was
4.6 per year (20 times lower).}  In addition, fainter potential companions
may have likewise been discovered with the various coronagraphs on board
GOES--19, SOHO, and STEREO-A between late March and late April (and
beyond), but their identities have not as yet been confirmed.

Before we describe our contribution to the problem of the orbital
motion and history of C/2026~A1, the primary objective of this
study, we briefly summarize the progress that has been reached
along these lines up to this point in time and what does it
allow us to conclude about the comet's distant companions, our
secondary objective.

\begin{table*}[ht] % Table 1
\vspace{0.15cm}
\hspace{-0.25cm}
\centerline{
\scalebox{1}{
\includegraphics{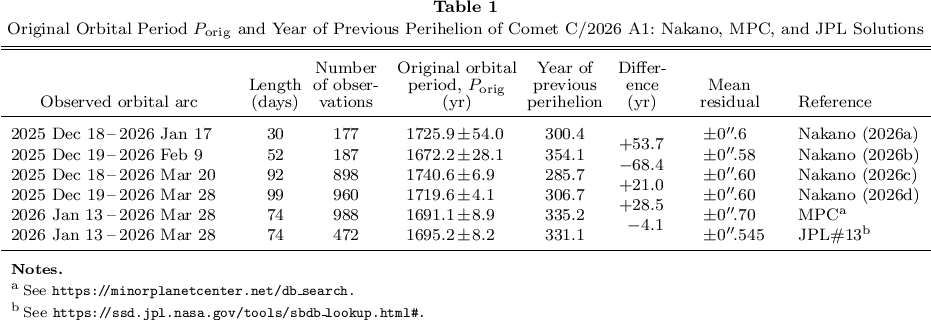}}}
\vspace{0.53cm}
\end{table*}

\section{Preliminary Orbits of C/2026 A1 and\\Distant Companions}
% Section 2
%
The most unexpected result of the early orbital computations
for comet C/2026~A1 --- all dealing with strictly gravitational
solutions and performed independently by S.~Nakano of the Central
Bureau of Astronomical Telegrams, by the staff of the Minor
Planet Center (MPC), and by the Jet Propulsion Laboratory's
(JPL) Solar System Dynamics group --- was the original orbital
period, $P_{\rm orig}$,\footnote{The original orbital period,
$P_{\rm orig}$, is referred to the barycenter of the Solar
System at an osculation epoch before the motion of a comet,
which is on its way to perihelion, begins to be discernibly
perturbed by the planets; $P_{\rm orig}$ closely approximates
the span of time from the previous perihelion passage to the
current one.} of 1664 to 1688~years, approximately twice~as~long
as a Kreutz sungrazer's average orbital period determined in
the past (Section~3.1).  The exceptionally long period implied
that the comet's previous~peri\-helion had been reached between
the years AD~338 and 362, although because of the short orbital
arcs available (a maximum of 52~days) these results were plagued
by high uncertainty, from $\pm$28~years to $\pm$518~years.

As the observing campaign continued, new sets of orbital elements
were computed; none of them incorporated nongravitational terms
in the equations of motion.  Even though the formal error of the
orbital period was being gradually reduced, a disconcerting problem
emerged, as $P_{\rm orig}$ turned out to depend rather strongly and
somewhat chaotically on the orbital arc employed.  This trend is
apparent from comparison of four orbital runs by Nakano (2026a,
2026b, 2026c, 2026d) and one each by MPC and JPL, all displayed in
Table~1.  Among the Nakano solutions the orbital period is seen
to increase by nearly 70~years, 2.5~times the higher and 10~times
the lower standard deviation, as the arc of used observations
extends from 52~days to 92~days; and then the period decreases by
more than 20~years, 3~times the higher and 5~times the lower
standard deviation, as the observed arc extends from 92~days to
99~days.  From a 74~days long arc the MPC solution provides a
period that is nearly 30~years shorter than Nakano's final orbit,
again more than 3~times the higher and 7~times the lower
standard deviation.  The only exception is the pair of periods
from the MPC and JPL orbits, both from the same orbital arc,
which agree with one another to better than one standard deviation.
But in general, the disparity between the standard deviations and
actual variations among the entries in Table~1 strongly suggests
that the comet's orbital motion has been subjected to forces whose
effects have not been properly accounted for in the tabulated
solutions and which need attention.

Even though a gravitational solution may appear to adequately fit
the astrometric observations, showing a distribution of positional
residuals with no alarming systematic trends, we maintain that
proper weighting of the employed data and, possibly, incorporation of
fitting nongravitational terms into the equations of motion should offer
the avenue for orbit improvement that is preferable to application
of an overstretched gravitational solution with an arbitrary cutoff
for the residuals subjected to cursory inspection.

Detection of likely distant companions of C/2026~A1 suggests that
the observed comet was not at all identical with the object that
separated from its parent in the 4th century --- the problem
addressed in the following section.  Although Kreutz (1901)
believed that the sungrazing comets break up only at perihelion,
the strongly episodic distribution of the SOHO dwarf sungrazers
has offered compelling evidence that nontidal fragmentation has
been taking place along the entire orbit (Sekanina 2000).  Distant
companions of C/2026~A1 should be products of exactly this kind of
event, and any of the comet's available sets of orbital elements
can be used to provide ballpark estimates for locations of such
fragmentation episodes.

Let us assume that a fragment detached, with a particular separation
velocity vector {\boldmath $V_{\bf sep}$}, from its parent comet at time
$t_{\rm frg}$, when the comet's \mbox{heliocentric} distance was equal to
$r_{\rm frg}$.  We measure $t_{\rm frg}$ from the time~of the comet's
subsequent perihelion passage, $t_\pi$, and write \mbox{$\Delta t_{\rm frg}
= t_{\rm frg} \!-\! t_\pi < 0$}.  Given this separation time and~the separation
velocity vector, we can readily determine the fragment's perihelion time,
$t_\pi^\prime$, which is relative to the comet's perihelion time measured
by \mbox{$\Delta t_\pi = t_\pi^\prime \!-\! t_\pi$} and is an observed
quantity.

\begin{table}[ht] % Table 2
\vspace{0.15cm}
\hspace{-0.21cm}
\centerline{
\scalebox{1}{
\includegraphics{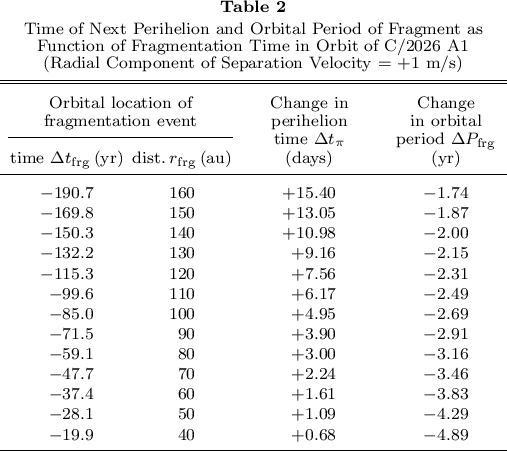}}}
\vspace{0.6cm}
\end{table}

Computations show that a fragment's perihelion time is affected primarily by
the component of the separation velocity vector that in the rotating comet-Sun
coordinate system points in the radial direction, {\boldmath $V_{\bf R}$}.
In Table~2 the perihelion time of a fragment, $\Delta t_\pi$, is presented
as a function of the fragmentation time, $\Delta t_{\rm frg}$, and
heliocentric{\vspace{-0.005cm}} distance, $r_{\rm frg}$, for a radial
separation velocity of +1~m/s {\vspace{-0.005cm}}away from the Sun,
causing the fragment to arrive at perihelion always {\it later\/}
than the comet.

The fragment's perihelion time depends on the separation velocity almost
perfectly linearly, so that the tabulated values of $\Delta t_\pi$ can
readily be scaled for any other low value of the velocity.  Unfortunately,
given that the velocity is unknown, the exact time of fragmentation
cannot be determined.  However, for the velocity in a general
submeter-per-second to meter-per-second range, the brightest companion
that followed the comet by the estimated 4.4~days was likely to have
separated some 50 to 130 years before perihelion at a heliocentric
distance around 70 to 130 au.

\begin{table}[b] % Table 3
\vspace{0.6cm}
\hspace{-0.21cm}
\centerline{
\scalebox{1}{
\includegraphics{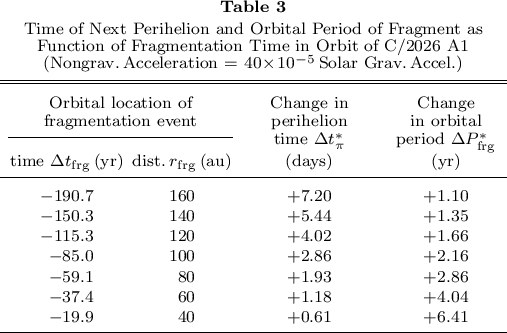}}}
\vspace{0cm}
\end{table}

Another effect of the separation velocity on the companion was that its
orbital period, $P_{\rm frg}$, got changed relative to that of the
comet's initial period, $P_0$.   In Table~2 the change is measured by
\mbox{$\Delta P_{\rm frg} = P_{\rm frg} \!-\! P_0$}.  Specifically,
the period of the companion arriving at perihelion later than the comet
was {\it shortened\/} typically by several years.  And because of the
conservation of momentum law, the comet's perihelion time and orbital
period changed in the opposite direction than the companion's, in inverse
proportion to their masses.  Thus, the comet arrived a little {\it
sooner\/} than it would have if there were no fragmentation, and its
orbital period {\it before\/} the event was a little {\it shorter\/} than
the one measured by the astrometric observations.  Of course, arguments
of this kind apply separately to each of the breakup episodes that the
comet appears to have undergone.

A companion's motion relative to the comet was also affected by a
differential, outgassing-driven nongravitational acceleration.  One
can argue that this effect was generally of lesser consequence than
the separation velocity, because in the opposite case no companion
would have been observed to precede the comet.  Nonetheless, a minor
but nontrivial contribution from the nongravitational acceleration
was unavoidable and should be accounted for.  The numbers in Table~3,
whose format is that of Table~2, refer to a relatively high acceleration
of \mbox{$40 \!\times\! 10^{-5}$} the solar gravitational acceleration,
equivalent to the standard parameter of \mbox{$A_1 = 11.8 \!\times\!
10^{-8}$\,au/day$^2$}.  For a given $\Delta t_{\rm frg}$ the time delay
at perihelion, $\Delta t_\pi^\ast$, is approximately proportional to
the magnitude of the nongravitational acceleration.  The correction
$\Delta P_{\rm frg}^\ast$ to\vspace{-0.05cm} the orbital period of
the companion makes it always longer than the period of the comet.

The very existence of the distant companions has implications for the
orbital motion of C/2026~A1, which is affected, however slightly, by
their separation long before the comet's discovery.  Hence, no orbital
solution~based on the astrometric observations can offer the comet's
true perihelion time in the 4th century, before the orbit was altered
by the birth of the distant companions.

\section{Comet C/2026 A1 and the Architecture of\\the Kreutz System}
% Section 3
%
Orbital periods of the Kreutz sungrazers are notoriously hard
to determine.  This is particularly true about the objects that
were under observation only before, or only after, perihelion ---
C/2026~A1 being one of them.  The sungrazers whose orbital periods
are known to better than, say, $\pm$40~years can be counted on the
fingers of one hand.  Yet, the orbital period is an extremely
important parameter because its knowledge allows one to track the
history of the given sungrazer and facilitates the modeling of the
origin and orbital evolution of the entire Kreutz system.

The original orbital period, as defined in footnote~3, is a
reasonable approximation of the time elapsed between two consecutive
returns of a comet to perihelion, the current and the previous.
Before the arrival of C/2026~A1 the meager statistics suggested
that Kreutz comets complete one orbit about the Sun in 700--900~years.
These numbers were in line with the general belief, now
supported by compelling arguments, that the spectacular
Great March Comet of 1843 (C/1843~D1) previously appeared
as the brilliant, widely observed Great Comet of 1106
(X/1106~C1), which in turn was one of the daylight comets in
late AD~363, recorded by the Roman historian Ammianus Marcellinus,
which in turn was the return of Aristotle's comet of 372~BC following
its nontidal fragmentation.  This remarkable string of dates,
implying an average period of 738~years, with a standard deviation
of less than $\pm$5~years, has served as one of the cornerstones of
a recent contact-binary hypothesis of the Kreutz system, defining its
Population~I (Sekanina 2021 and following).  Similar sequences of
major sungrazers (and probable historical sungrazers) have described
Population~II and may describe other populations as well.

\begin{table*}[ht] % Table 4
\vspace{0.13cm}
\hspace{-0.21cm}
\centerline{
\scalebox{0.99}{
\includegraphics{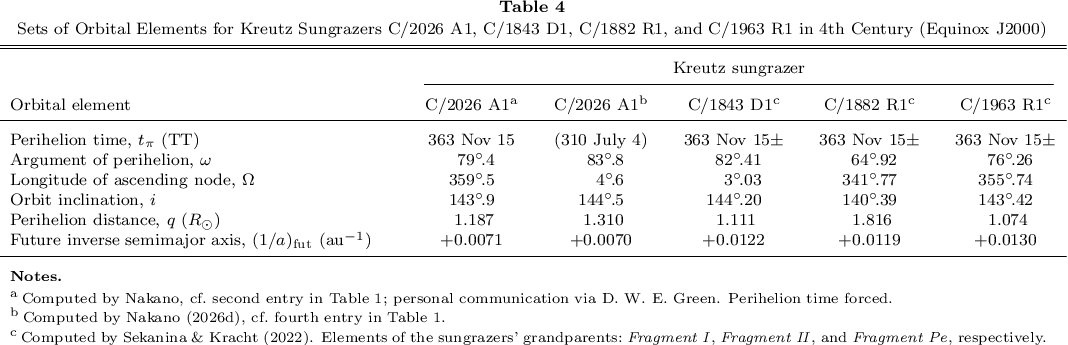}}}
\vspace{0.48cm}
\end{table*}

\subsection{The Problem of the Orbital Period of C/2026 A1} % Section 3.1
Given that the nucleus was small, subkilometer sized, and the previous
perihelion --- determined via the original barycentric orbit --- occurred
almost certainly in the 4th century, a reasonable conclusion in the context
of the contact-binary hypothesis was that comet C/2026~A1 represented
(or was part of) a minor, outlying fragment of one of the bright daylight
comets recorded in late 363.  Accordingly, as argued elsewhere (Sekanina
2026a), the apparent upshot of the surprisingly long orbital period was
that the comet portrayed the only known second-generation fragment of 
Aristotle's comet that arrived at perihelion later than in the 12th
century.  Yet, the disconcertingly large disagreement among the results
of the six gravitational solutions presented in Table~1 suggests that
the determination of the comet's orbital period has by no means been
straightforward and that the problem does deserve close attention.

An alternative and obviously more compelling line of attack is direct 
integration of the comet's 2026 orbit back to the 4th century's
perihelion, which affords one the complete set of six elements,
not only the perihelion time.  Compared with the outcome of direct
integration of the motions of three major sungrazers by Sekanina
\& Kracht (2022) in Table 4, comparison of Nakano's two orbital
solutions for comet C/2026~A1 does not at first sight contribute much
to the understanding of the comet's history, as the sets of elements
are rather discordant:\ they differ 47~years in the perihelion time,
over 5$^\circ$ in the nodal longitude, and one eighth of the Sun's
radius in the perihelion distance.  And even though there is
a fair degree of agreement between Nakano's second solution and
the orbit of the comet of 1843 in the argument of perihelion,
nodal longitude, and orbit inclination, the disparity in the perihelion
distance does not support the hypothesis of firsthand connection
between the two objects.

A seemingly extreme, yet insightful point that illustrates the
potentially enormous influence of the nongravitational acceleration
on the orbital period when the eccentricity is near unity follows
from straightforward comparison of a Kreutz sungrazer with a comet
arriving from the Oort cloud.  Nearly 50~years ago Marsden \&
\mbox{Sekanina} (1978) determined that, on account of a {\it neglected\/}
nongravitational acceleration, the computed orbit of an Oort reservoir
comet appeared the more hyperbolic the smaller was its perihelion
distance, $q$.  The inferred relationship between $1/q$ and the
original reciprocal semimajor axis, $(1/a)_{\rm orig}$, turned out
to be linear, so that an average Oort cloud comet of the perihelion
distance equaling that of C/2026~A1 would have had{\vspace{-0.023cm}}
\mbox{$(1/a)_{\rm orig} \simeq -0.004090$~au$^{-1}$} instead of
+0.000046~au$^{-1}$.  If C/2026~A1 were subjected to an effect of
this magnitude, its ``observed'' orbital period of 1719.6~years
(the fourth entry in Table~1) should be corrected to 855~years and
the comet should have arrived at perihelion in 1171, not in 306!
Of course, Kreutz sungrazers are much less active than Oort cloud
comets and their motions are, accordingly, subjected to much smaller
perturbations of this kind.  But even if the magnitude of the
nongravitational effect were a factor of 25 lower, it still would
have changed the orbital period by 60~years, to 1660~years.  The
perihelion would have been reached in 366, three years {\it later\/}
than the expected return.  An outgassing-driven nongravitational
force could indeed make a lot of difference.

\subsection{Weighting of Observations, and Virtual Clones} % Sextion 3.2
Every astrometric observation of a comet is burdened by an error, whose
magnitude depends on the appearance of the object (such as its brightness
or degree of condensation of its head), the experience of the observer,
his instrumentation (such as the scale and quality of imaging), observing
conditions (such as transparency of the atmosphere or the elevation above
the horizon), and the method of reduction.  The less favorable these
circumstances get, the less useful the astrometric observation is for
orbit determination and the larger residual it usually leaves in a
least-squares solution.

Various approaches are adopted to deal with such situations.  A
frequently used approach is to assign equal weights to all
observations, but adopt an arbitrary minimum residual limit to
eliminate inaccurate observations.  This may lead to the selection of
a restricted set of observations regarded as the most reliable, while
entire sets of observations made elsewhere are rejected.  In the case
of comets, often subjected to systematically acting nongravitational
forces, use of this procedure may lead to misinterpretion of trends
in the observations as evidence of poor-quality data.  In reality,
such trends, which are also responsible for a noticeable departure
of the distribution of positional residuals from a Gaussian law,
could indicate the existence of appreciable nongravitational forces,
as noted in Section~3.1.

\begin{table*}[ht] % Table 5
\vspace{0.13cm}
\hspace{-0.21cm}
\centerline{
\scalebox{0.975}{
\includegraphics{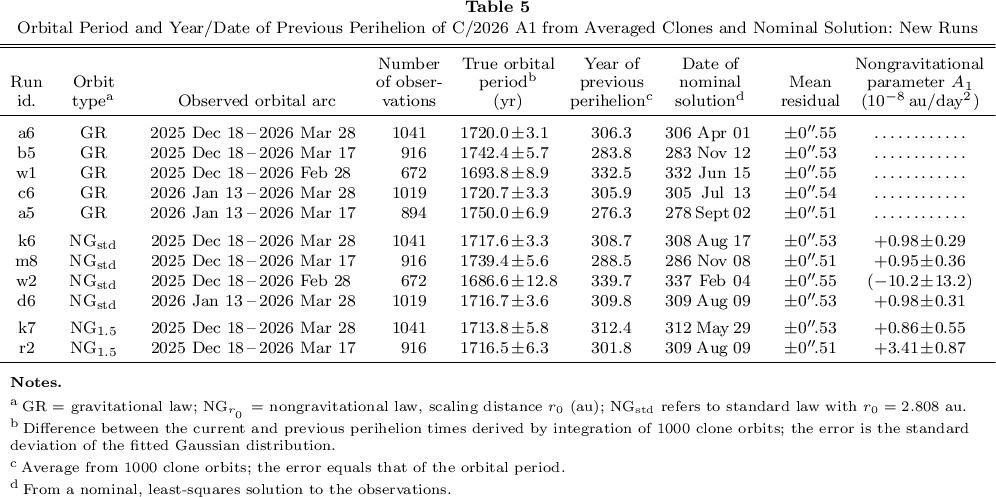}}}
\vspace{0.42cm}
\end{table*}

A derived set of orbital elements could be affected by such not
fully objective procedures.  In the interest of optimizing the
orbit-determination process, we employ a specific, elaborate method
of selecting and weighting cometary positional data.  The technique
was developed over a period of time by members of the Space Research
Centre of the Polish Academy of Sciences in Warsaw (Bielicki \&
Sitarski 1991) and was described in detail by Kr\'olikowska et
al.\ (2009), Kr\'olikowska \& Dybczy{\'n}ski (2010), and references
therein.

The employed method assigns a weight to each observation which is
a function of the residual and rejects a data point only when the
probability of its deviation from the mean is less than the inverse
of the number of observations (Bessel criterion) or the inverse of
twice the number of observations (Chaufenet criterion), depending
on the particular observational dataset under consideration.  In
the present study, the more restrictive Bessel criterion is applied.
In the iterative weighting procedure of the orbit determination
process, relative weights are assigned to datasets from individual
observatories, with the inverse of the weights normalized so that
their sum equals unity.  This means assigning relative weights
according to the actual scatter of the residuals with respect~to~the
computed orbit for each contributing observatory.

The quality of orbit fitting is judged by three conditions:\ the
rms deviation, the systematic trends in the residuals, and the
compatibility of the distribution of residuals with the Gaussian
distribution.  The past results obtained in this manner were in
excellent agreement with those available from the MPC and JPL
Small-Body Database (Kr\'olikowska \& Dones 2023), and have
routinely been used in the CODE catalogue (Dybczy{\'n}ski \&
Kr\'olikowska 2025).

In this procedure, departures of residuals from the Gaussian distribution
obtained for a gravitational orbit are regarded primarily as
evidence of nongravitational effects.  If such deviations are
statistically significant, the remedy is to compute a nongravitational
orbit.  If this does not appear to help, the culprit could be the adopted
law for the nongravitational acceleration that does not adequately describe
the comet's physical~behav\-ior.~A variety of nongravitational laws should
then~be~tested.

Sitarski's (1998) method of random orbit selection is used to generate
a body of 1000 virtual clones from each nominal orbital solution.  The
procedure produces Gaussian distributions in the six-to-ten-dimensional
space of orbital elements and, if included, nongravitational parameters.
The clones' orbits then serve to determine the most probable time of
the previous perihelion (Table~5).\,\,\,\,

\subsection{Gravitational Solutions with Weighted Observations} % Section 3.3
\vspace{-0.04cm}
One of our concerns is the discrepancy in Table~1 of some 25--30~years in
the original orbital period between Nakano's gravitational solution based
on an orbital arc of 2025 December~19--2026~March~28 (entry~4) on the one
hand, and the MPC and JPL gravitational solutions based solely on the
post-discovery arc (from 2026 January 13 to 2026 March 28; entries~5 and 6,
respectively) on the other hand.  The issue is whether the small number
of pre-discovery observations from 2025 December~18 to 2026~January~10, a
little more than 2~percent of the total number of observations, could make
such a large difference, a priori judged as highly unlikely.

To address this issue, we first derive a pair of gravitational orbits,
with and without the pre-discovery observations.  The results, listed
in Table~5 as {\it Runs~a6\/} and {\it c6\/}, respectively, leave no
doubt whatsoever that in terms of the orbital period, it makes
practically no difference whether or not the pre-discovery astrometry 
is included.  This conclusion is further supported by {\it Runs~b5\/}
and {\it a5\/}, based on orbital arcs ending on March~17 rather than
March~28 (see below), and by a pair of nongravitational solutions
(Section~3.4).  Hence, the apparent disparity in Table~1 is caused
by something else.  Given that the compatible MPC and JPL values of
the original orbital period were computed from different sets of
observations, with use of different osculation epochs, we see no
obvious source of the disparity.  The result of Run~a6 agrees
exceptionally well with Nakano's (2026d) orbital period.\,\,\, 
 
The distributions of residuals over a period of time from 2025
December~18 to 2026 March~28 reveal that the observations made
later than March~17 render them non-Gaussian, with the peak
deviating appreciably from the optimum position.  This anomaly,
which makes the results of Runs~a6 and c6 highly questionable,
is mitigated in {\it Runs b5\/} and {\it a5\/}, based on orbital
arcs terminating on March~17.  We note that the orbital period
is now predicted to have lengthened compared to those from Runs~a6
and c6 by \mbox{20--30}~years, a difference greatly exceeding
the standard deviation.  Runs~b5 and a5 thus imply that the
previous perihelion occurred in the late 3rd century, at least
80~years before the expected time, a rather disturbing outcome.
However, it is compatible with Nakano's result for the nearly
identical orbital arc (see entry~3 of Table~1), suggesting that
the predicted orbital period depends on how close to the Sun do
the observations used in the computations extend.

Guided by this experience, our next step is a somewhat arbitrary
reduction of the observed arc's length by terminating it at the end
of February.  A gravitational solution based on the observations
from 2025 December~18 till 2026 February~28 is listed in Table~5
as {\it Run w1\/}.  The derived orbital period is nearly 50~years {\it
shorter\/} than that from Run~b5 and the previous perihelion is now
predicted to have taken place in the 330s.

\begin{figure}[b] % Figure 1
\vspace{0.55cm}
\hspace{-0.19cm}
\centerline{
\scalebox{0.7}{
\includegraphics{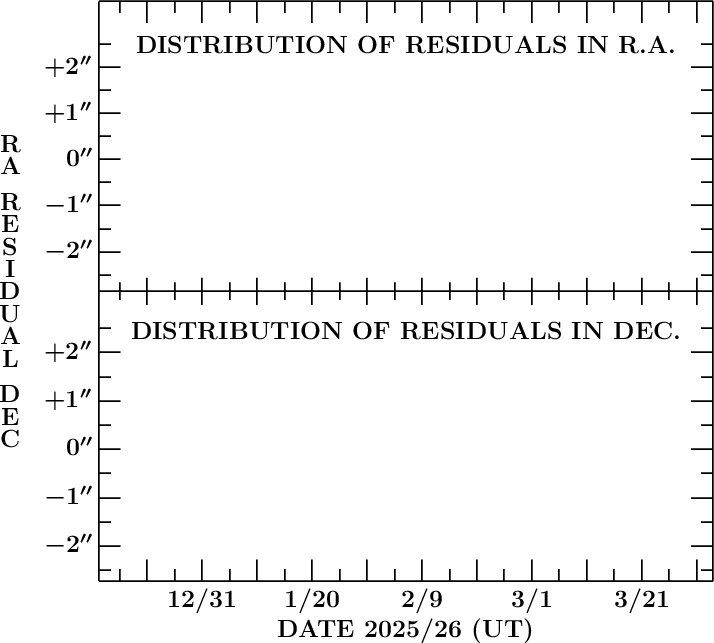}}}

\vspace{-6.915cm}
\hspace{0.45cm}
\centerline{
\scalebox{0.3}{
\includegraphics{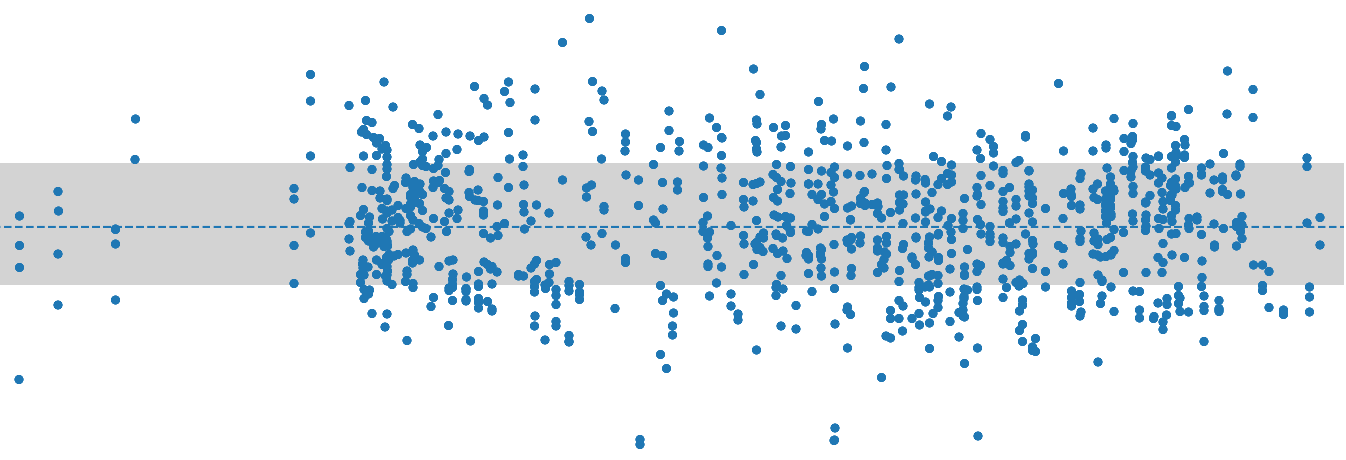}}}

\vspace{1.11cm}
\hspace{0.45cm}
\centerline{
\scalebox{0.3}{
\includegraphics{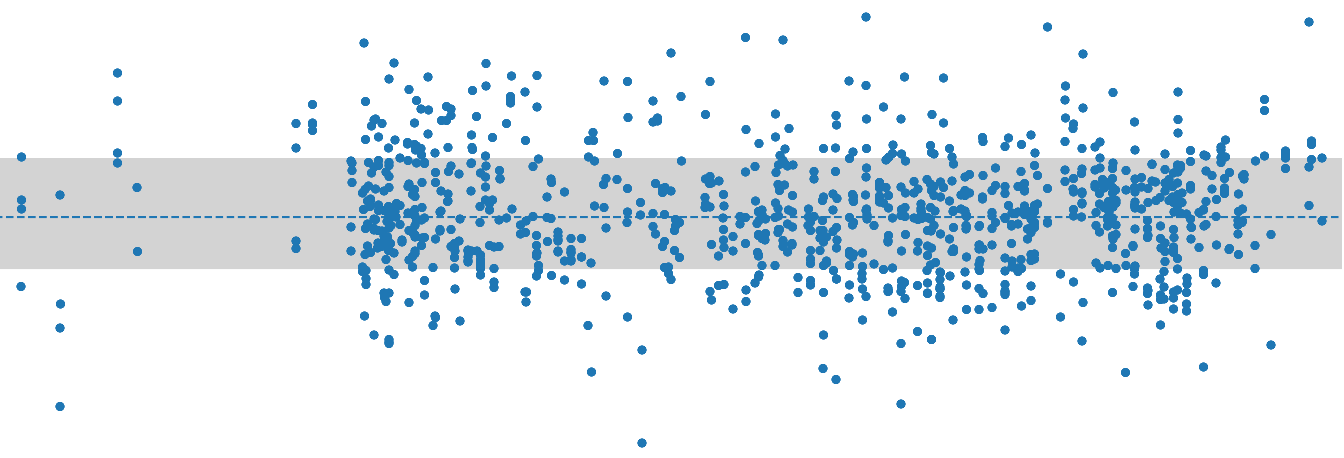}}}		
\vspace{0.7cm}
\caption{Distribution of positional residuals from 1041 astrometric
observations of comet C/2026~A1 between 2025 December~18 and 2026 March~28
displayed by Run~a6, a gravitational solution, in right ascension (top)
and declination.  The observations before 2026 January~13 are
pre-discovery ones.  The grey bands show the mean residual.  Note that
the distribution of points is not perfectly random along the
vertical axis, revealing slight systematic trends.}
\vspace{-0.05cm}
\end{figure}

Even though Runs b5 and w1 are superior to Run~a6 in terms of the
distribution of residuals, the strong disagreement in the orbital
period still is a problem.  Our limited experimentation with the
gravitational solutions appears to suggest that, at least for
solutions from time intervals terminating not later than March~17,
the earlier is the date of the last observation used in the
computations, the shorter is the derived orbital period, thereby
moving the 4th-century perihelion time forward.

\subsection{Nongravitational Solutions with\\Weighted Observations}
% Section 3.4
%
We find that the least controversial nongravitational orbits are
those in which the coefficient $A_1$ of the radial component is the
only nongravitational parameter solved for, i.e., \mbox{$A_2 = A_3 =
0$}.  As a rule, such runs result in \mbox{$A_1 > 0$}, the acceleration
pointing away from the Sun, but {\it Run w2\/} is an exception.  Most
test runs in which we have solved for both $A_1$ and $A_2$ offer
meaningless results:\ either \mbox{$A_1 < 0$} or $A_2$ is indeterminate
(the error exceeding the absolute value of $A_2$); these runs should be
ignored.

Most runs use Marsden et al.'s (1973) standard~non\-gravitational law
$g(r)$, whose scaling distance \mbox{$r_0 \simeq 2.8$ au} is diagnostic of
the sublimation of water-ice.  A few runs employ instead a substantially
lower scaling distance of 1.5~au (keeping the other parameters
unchanged), appropriate for the sublimation of more refractory
substances, such as formic acid (HCOOH) or form\-amide (NH$_2$CHO).

The nongravitational runs listed in Table 5 generally support our conclusions
based on the gravitational runs.  In particular, {\it Runs~k6\/} and
{\it d6\/} confirm that incorporation of the pre-discovery observations,
while improving the determinacy, have essentially no effect on the
derived orbital period (Section~3.3).  As expected, its values are
--- particularly for the nonstandard scaling distance --- a little
lower than displayed by the gravitational solution, and the errors
higher.  The error of $A_1$ in {\it Run~k7\/} is well over 50~percent
and therefore unacceptable.  Most importantly, comparison of {\it
Runs~m8\/}, {\it w2\/}, and others shows that the disappointing lack
of agreement of the derived values of the orbital period (Section~3.3)
persists among the nongravitational runs as well.

\subsection{Critical Appraisal of the Orbital Solutions} % Section 3.5
In-depth analysis of astrometric residuals from the gravitational
solutions listed in Table~5 (Runs~a6, b5, w1, c6, and a5),
covering observational arcs between approximately two
months (Run~w1) and more than three months (Run~a6), reveals
a systematic trend, the nearly constant mean residual
($\pm$0$^{\prime\prime\!}$.51 to $\pm$0$^{\prime\prime\!}$.55)
and application of a robust iterative data weighting and rejection
procedure notwithstanding.  An example, the distribution of
residuals from Run~a6, is displayed in Figure~1.

As addressed in some detail in Section~4, the comet's complex
activity makes major deviations from a simple non-gravitational
law essentially inevitable.  This conclusion is supported by
our finding that orbital solutions based on the standard or
similar nongravitational law (Runs~k6, m8, w2, d6, k7, and r2
in Table~5) do not generally do better than the equivalent
gravitational solutions.

Over all, the reduction procedure employed did typically
produce near-Gaussian distributions of residuals when the
adopted dynamical model adequately represented the comet's
motion.  Importantly, the weighting and rejection scheme
has always been applied consistently across all solutions
and has not introduced artificial asymmetries or distortion
of the distributions of residuals.

For the shortest solution (Run~w1), the residuals in right
ascension are fully consistent with a single Gaussian
population, with negligible skewness and kurtosis and
they show no signs of non-Gaussian features. The
residuals in declination show only moderate departures
from the Gaussian law.  However, the more extended
observational arcs (Runs~a6, b5, and c6) display systematic
degradation of the distribution of residuals in right
ascension.  The effect is characterized primarily by
increasingly negative kurtosis (platykurtic behavior)
while maintaining very small skewness.  At the same
time, the residuals in declination remain comparatively
stable and, for some solutions, even move closer to the
Gaussian law.  Thus, the deterioration is highly
asymmetric between the two coordinates.  Crucially,
this transition is accompanied by no obvious temporal
feature in the residuals, nor by a substantial increase
in the mean error.  Plotted as a function of time, the
residuals do not reveal evidence of clear discontinuities,
trends, or sudden events.  Moreover, the described effect
persists independently of whether or not the pre-discovery
observations are included, suggesting that the observed
behavior is not caused by a specific subset of astrometric
data.

Among the nongravitational runs, dynamically meaningful
two-parameter solutions with both \mbox{$A_1 > 0$} and
\mbox{$|A_2| < A_1$} can only be obtained for the shorter
arcs (before approximately 2026 Mar 17).  In contrast,
after extending the arc beyond this date, stable
solutions generally require restricting the model to
$A_1$ only.  Solving for $A_2$ leads to unstable or physically
problematic solutions, including negative $A_1$ values.
This behavior strongly suggests a growing degeneracy
between orbital and nongravitational parameters as the
arclength increases.  At the same time, introduction
of nongravitational models based on canonical, other
$g(r)$-like laws produces at best only marginal
improvements in the mean residual and does not
restore a Gaussian distribution of residuals.  Taken
together, these results suggest that the observed
non-Gaussian feature is unlikely to originate from
improper weighting, inadequate rejection of outliers,
or contamination of the astrometric dataset. The
iterative weighting procedure already accounts for
heterogeneous observational quality, while the Bessel
rejection criterion effectively removes statistically
inconsistent measurements independently in either
coordinate.  If poor-quality observations were the
dominant cause, one would expect stronger degradation
in terms of the mean residual, identifiable temporal
features, or clear improvement after removing specific
data subsets.  None of these effects is observed.
Instead, the emerging picture is more consistent with
a transition from a regime dominated primarily by
measurement noise to one where structural model
effects and parameter degeneracies become significant.
The distributions of residuals appear to reflect not
only observational uncertainties but also limitations
in the identifiability of the dynamical model over a
relatively short-to-medium arc for a physically complex
sungrazing object.  In particular, the instability of
the two-parameter nongravitational solutions after
mid-March may indicate that the symmetric $g(r)$-type
formalism is no longer adequate to represent the
comet’s true activity, which could have become
episodic, asymmetric, or otherwise non-stationary
as the comet was approaching perihelion and eventual
disruption.

Consequently, the observed degradation of Gaussian
behavior should not be interpreted as a justification
for aggressive rejection of observations.  Removing
large fractions of data could artificially improve
the apparent statistical properties of the residuals
while simultaneously reducing the dynamical information
content of the orbital solution.  Evidence instead
supports the interpretation that the observed effects
originate primarily from limitations of the adopted
dynamical description and increasing parameter
correlations, rather than from deficiencies in the
astrometric dataset itself.

\section{Comet's Activity and Its Potential Effects\\on the
 Orbital Motion} % Section 4
Although an obvious choice, we have not used activity as a
gauge to find the end dates of the orbital arcs for deriving
the solutions presented in Table~5.  As a remedy, we next inspect
the light curve, the only characteristic of the comet available
to us at this time that shows temporal variations in the rate
of outgassing from the nucleus.  The goal is to search for
changes that could suggest potentially detectable effects on
the orbital motion.  Closer to the Sun (from the beginning
of March on) the comet's fairly irregular rate of brightening
has extensively been documented (e.g., Green 2026b), but our
interest is focused on the early period of time, when the comet
was at heliocentric distances greater than 1~au.

\begin{figure}[b] % Figure 2
\vspace{0.75cm}
\hspace{-0.21cm}
\centerline{
\scalebox{0.65}{
\includegraphics{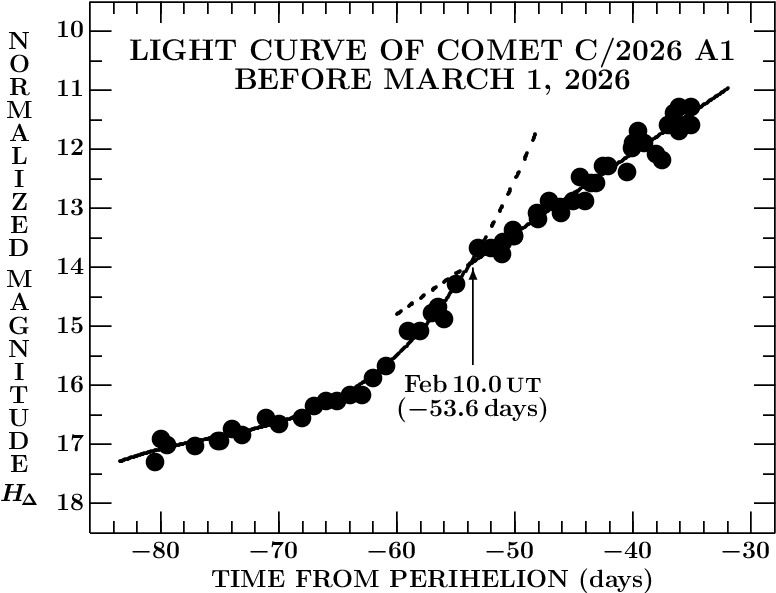}}}
% \vspace{-0.1cm}
\caption{Light curve of comet C/2026 A1 between January~14 and
 February~28, 2026, or 80 to 35~days before perihelion.  The
 total CCD~magnitude is normalized to unit geocentric distance
 and zero phase angle, employing a phase law for dust-poor comets
 proposed by Marcus~(2007).  The light curve has a deflection~point
 on February~9--12, when an early rate of rapid brightening~abruptly
 changes into a much slower and noisier climb.  The data points are
 CCD observations by T.\ Lovejoy with a 25-cm f/4 reflector, 32-cm
 f/9 Cassegrain, 43-cm f/7 Cassegrain, and 51-cm f/7 Casse\-grain;
 M.\ Ma\v{s}ek with a 30-cm f/6.8 Cassegrain;~M.~\mbox{Mattiazzo}~with
 a 28-cm f/2 Schmidt; and A.\ R.\ Pearce with a 35-cm f/6
 Schmidt-Cassegrain and 43-cm f/7 Schmidt-Cassegrain.~These~\mbox{magnitude}
 observations are available from the Comet Observation~Database (COBS)
 website maintained by the {\it \v{C}rni~Vrh~\mbox{Observatory}\/},
 Slovenia ({\tt https:/$\!$/cobs.si/analysis)}.}
\vspace{-0.05cm}
\end{figure}
\begin{table*}[t] % Table 6
\vspace{0.13cm}
\hspace{-0.21cm}
\centerline{
\scalebox{1}{
\includegraphics{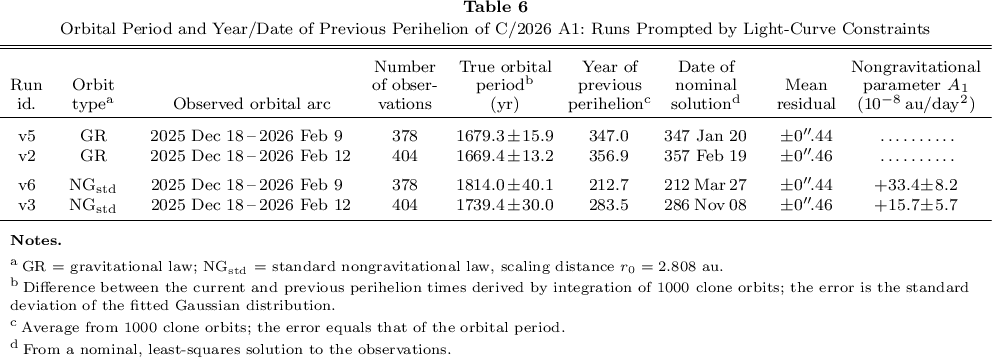}}}
\vspace{0.6cm}
\end{table*}

Figure 2 displays the comet's light curve between discovery and
the end of February.  Plotted are total CCD magnitudes, extracted
from the list of observations reported to the Comet Observation
\vspace{-0.065cm}Database (COBS) website maintained by the {\it
\v{C}rni Vrh Observatory\/}, Slovenia.  Each apparent magnitude
has been normalized to 1~au from the Earth and zero phase angle.
Although obtained by four observers (T.\ Lovejoy, M.\ Ma\v{s}ek,
M.\ Mattiazzo, and A.\ R.\ Pierce; see the caption to Figure~2 for
details), the chosen magnitudes ($\sim$40~percent of the total
listed in COBS over the selected time range) are remarkably
consistent with no personal corrections applied.  The phase
correction has been computed from a law proposed by Marcus (2007)
for dust-poor comets, a choice consistent with high negative
$V\!-\!R$ color indices reported at these times (see the COBS
catalogue).

In the 45-day window in Figure 2, the light curve consists of two very
different parts.  The first part, extending over a period from January~14
to February~9--12 (80 to \mbox{51--54}~days before perihelion), the
comet brightened steadily at an accelerating rate, showing hardly any
short-term fluctuations.\footnote{From Figure 2 it follows that the
end date of this phase of the light curve is uncertain to a few days.
The figure shows one possible fit to the reported magnitude observations,
but equally successful fits could be achieved, if a few additional data
points on the days following February~10.0~UT are linked with the earlier
ones.  We suggest that February~\mbox{9--12} is a reasonable estimate
for the time of the anomalous feature --- the transition time between
the two segments of the light curve --- because of data noise.}
Shortly before mid-February a new phase of light evolution set rather
abruptly in.  The comet's rapid expansion of activity ceased, as if
one of emission sources on the nucleus was running out of gas or its
production was in a slump, and the intrinsic brightness began to
follow a pattern of more modest growth with gradually augmenting
interim variations superimposed, as the heliocentric distance continued
to drop.  As the anomaly involved no flare-up, it is unlikely that it
was associated with a fragmentation event, yet it certainly could affect
the comet's orbital motion.

\section{Principal Set of Orbital Elements of \mbox{C$\:\!\!$/2026} A1}
% Section 5
%
Regardless of whether the remarkable light-curve feature of
February~9--12 indeed was a turning point that marked the emergence
of measurable perturbations of the comet's motion due to erratic
activity, orbital solutions with these end dates are undoubtedly the
best that we can muster to approximate the comet's ``unadulterated''
motion, serving in the following as a means for ultimately selecting
the {\small \bf principal set of orbital elements\/}.

The two options with the end dates three days apart also provide a
convenient check on the level of noise that limits the accuracy with
which the principal orbit could in fact be determined.  One source
of this noise stems from the precariously short length of the observed
arc, which now amounts to less than two months; the other is associated
with the response of the least-squares routine to the more than two
dozen additional observations in the longer arc.

The two gravitational solutions, derived from the observations distributed
along the orbital arcs extending from 2025 December~18 through 2026
February~9 or 12, appear as, respectively, {\it Runs~v5\/} and {\it
v2\/}, in Table~6.~We also derive equivalent nongravitational
solutions, using the standard $g(r)$ law, and these are tabulated
as,~respec\-tively, {\it Runs~v6\/} and {\it v3\/}.

The most important results are an orbital period and date of the
previous perihelion, which are very different from the numbers in
Table~5.  The previous perihelion~is now~predicted to have taken place
in a range of AD~347 to 357.  The span of 10~years is well within
1$\sigma$, which is approximately $\pm$15~years, representing
effectively the uncertainty with which the time of the previous
perihelion is determined.  We note that from Run~v5 this time~is
just about 1$\sigma$ from the time of appearance of the daylight
comets in AD~363.  The agreement is much better, about 0.5$\sigma$,
for Run~v2.  Another remarkable result is~a significantly lower
mean residual compared to those from the orbital solutions in
Table~5.

The two nongravitational solutions listed in Table~6, {\it Runs~v6\/}
and {\it v3\/}, are meaningless.  They utterly fail to improve
the mean residual, yield very discordant and improbable orbital
periods and times of the previous perihelion, and imply a
nongravitational acceleration more than one order of magnitude
higher than follows from the equivalent solutions in Table~5.
In addition, the results are burdened by high uncertainties, and
the general impression is that the orbital arc is too short to
allow meaningful nongravitational solutions.

\begin{table*}[t] % Table 7
\vspace{0.14cm}
\hspace{-0.21cm}
\centerline{
\scalebox{1}{
\includegraphics{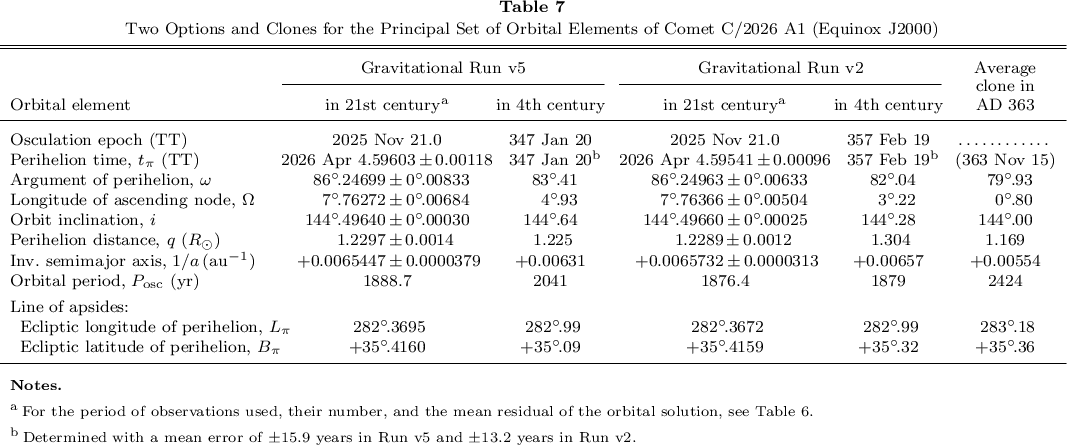}}}
\vspace{0.7cm}
\end{table*}

Table 7 presents complete sets of orbital elements from the two
gravitational solutions in Table~6, Runs~v5 and v2, which offer
our best candidates for the principal~set. By sheer chance, the
observed orbital arc essentially coincides with that of Nakano's
second solution in Table~1, but we now have more than twice as
many observations available as Nakano had at the time, which
substantially improves the determinacy of our results.  While
the time for the 4th-century perihelion predicted from either of
our runs agrees with Nakano's value to better than eight years,
our error bar is nearly a factor of two smaller.

We keep in mind that the tabulated solutions still do not provide
a truly realistic estimate for the length of the period of time
between the perihelion passages in the 4th and 21st centuries,
because the nontrivial orbital perturbations triggered in the
course of the fragmentation event(s) that gave rise to distant
companions remain unaccounted for, an issue addressed in Section~6.

\section{Estimating the Effects of Separation of\\Companions
 on the Comet's Motion} % Section 6
At present we do not know the total number of dwarf sungrazers
that accompanied C/2026~A1 on its journey to perihelion.  For
the purpose of the exercise presented below we only consider the
three bright, distant companions, discussed in Section~2.  With
the comet itself, we thus carry on with four fragments, assigning
designations customary for split comets:\ in the order in which
they would have passed perihelion (if they had survived), we refer
to them, respectively, as A (passing about 4.3~days before the
comet), B (the comet), C (passing about 4.4~days after the comet),
and D (passing about 3.4~days after C).  The task is to crudely
estimate, on a number of assumptions, the effects of separation
of A, C, and D on the perihelion time and orbital period of B
(i.e., the comet itself), not to examine them in detail, which
is not at present possible.

When there are more than one companion, as in this case, a major issue
is whether they are products of one or more fragmentation events.
Because of the very short observed orbital arcs of the companions
to C/2026~A1, the answer to this question will never be known with
certainty.  In the following we assume that all fragments were born
simultaneously, because that is the easiest case to handle.  We
further assume that the fragments separated from one another along
the radius vector and that their masses in units of the mass of the
original comet (before it fragmented) were ${\cal M}_{_{\rm A}}$,
\ldots, ${\cal M}_{_{\rm D}}$:
\begin{equation}
{\cal M}_{_{\rm A}} + {\cal M}_{_{\rm B}} + {\cal M}_{_{\rm C}}
 + {\cal M}_{_{\rm D}} = 1,
\end{equation}
where \mbox{${\cal M}_{\rm x} \ll {\cal M}_{_{\rm B}}$} \mbox{{\footnotesize
(X} = {\footnotesize A, C, D)}}, and therefore \mbox{${\cal M}_{_{\rm B}}
\simeq 1$}.  If the radial component of the separation velocity of
fragment X is $U_{\rm x}$, the conservation-of-momentum
law can under the adopted conditions be written as
\begin{equation}
{\cal M}_{_{\rm A}} U_{_{\rm A}} + {\cal M}_{_{\rm B}} U_{_{\rm B}} +
 {\cal M}_{_{\rm C}} U_{_{\rm C}} + {\cal M}_{_{\rm D}} U_{_{\rm D}} = 0.
\end{equation}
Since we already noted in Section 2 that $\Delta t_\pi$, the change
in the perihelion time imposed on a fragment~by~a~break\-up event,
varies as the radial component of the separation velocity, we replace
$U_{\rm x}$ with $(\!\Delta t_\pi\!)_{\rm x}$, the change in the
perihelion time of fragment~X (relative to the original comet),
getting for the magnitude of the effect of the nuclear fragmentation
on the perihelion time of the observed comet:
\begin{equation}
(\!\Delta t_\pi\!)_{_{\rm B}} = -\frac{\sum_{\rm (x)} {\cal M}_{\rm x}
 (\!\Delta t_\pi\!)_{\rm x}}{1 \!-\! \sum_{\rm (x)} {\cal M}_{\rm x}},
\end{equation}
where the summations are carried out (both here and below) over the three
companions, A, C, and D.  If these objects were subjected to no nongravitational
accelerations, Equation~(3) could serve as the starting formula for estimating
$(\!\Delta t_\pi\!)_{_{\rm B}}$.  In practice, we need to express
$(\!\Delta t_\pi\!)_{_{\rm B}}$ in terms of the observed relative perihelion
times of the companions, $(\!\Delta \tau_\pi\!)_{\rm x}$, which deviate
from $(\!\Delta t_\pi\!)_{\rm x}$~in that they are measured relative to the
motion of the {\it observed\/} comet (fragment~B), rather than the motion
of the original, 4th-century comet; and they include the contributions
$(\!\Delta t_\pi^\ast\!)_{\rm x}$ from the companions' differential
nongravitational accelerations (cf.\ Table~3), which are also reckoned
relative to the observed comet:
\begin{equation}
(\!\Delta \tau_\pi\!)_{\rm x} = (\!\Delta t_\pi\!)_{\rm x}
 - (\!\Delta t_\pi\!)_{_{\rm B}} + (\!\Delta t_\pi^\ast\!)_{\rm x}.
\end{equation}
Inserting for $(\!\Delta t_\pi\!)_{\rm x}$ from Equation~(4), Equation~(3)
becomes
\begin{equation}
(\!\Delta t_\pi\!)_{_{\rm B}} = -\!\sum_{\rm (x)} {\cal M}_{\rm x} \!\left[
 \:\!\! (\!\Delta \tau_\pi\!)_{\rm x} - (\!\Delta t_\pi^\ast\!)_{\rm x} 
 \raisebox{0ex}[1.1ex][0.8ex]{} \:\!\!\right] \! .
\end{equation}

Given~that\,\mbox{$(\!\Delta \tau_\pi\!)_{_{\rm A}} \!=\! -4.3$
days},\,\mbox{$(\!\Delta \tau_\pi\!)_{_{\rm C}} \!=\! +4.4$ days},\,and
\mbox{$(\!\Delta \tau_\pi\!)_{_{\rm D}} \!=\! +7.8$ days}, $(\!\Delta
t_\pi\!)_{_{\rm B}}$~can~be~evaluated~from~con\-dition~(5) for
assumed values of ${\cal M}_{_{\rm A}}$, ${\cal M}_{_{\rm C}}$,
${\cal M}_{_{\rm D}}$,~$(\!\Delta t_\pi^\ast\!)_{_{\rm A}}$,
$(\!\Delta t_\pi^\ast\!)_{_{\rm C}}$, and $(\!\Delta t_\pi^\ast\!)_{_{\rm
D}}$.  For example, if we adopt \mbox{${\cal M}_{_{\rm C}} \!=\! 0.05$},
\mbox{${\cal M}_{_{\rm A}} \!=\! {\cal M}_{_{\rm D}} \!=\:\!\! 0.01$}, and
furthermore \mbox{$(\!\Delta t_\pi^\ast\!)_{_{\rm C}} \!=\:\!\! 0.4$ day}~and
\mbox{$(\!\Delta t_\pi^\ast\!)_{_{\rm A}}\!=\:\!\!(\!\Delta t_\pi^\ast\!)_{_{\rm
D}} \!=\:\!\! 0.85$ day}, we find\footnote{The masses chosen for the three
companions are entirely~arbitrary, but the nongravitational acceleration's
effects~on~the~perihelion times of the companions are partially constrained,
as they vary inversely with a cube root of the mass.  Thus, given that the
computations in Table~3 apply~to~\mbox{$A_1 \approx 12$} in units of
$10^{-8}$\,au/day$^2$ and the observed comet (fragment~B) had $A_1$ near
unity, an estimated effect from a breakup\vspace{-0.055cm} at 100~au from
the Sun is 0.24~day.  For companion~C, $A_1$ is \mbox{$(0.93/0.05)^{1/3}
= 2.65$ times} greater and \mbox{$(\!\Delta t_\pi^\ast\!)_{_{\rm C}} = 0.64
\!-\! 0.24 = 0.40$~day} relative to the observed comet.  For companions A
and D the effect amounts~to~1.09$\:\!-\:\!$0.24 = 0.85~day.}%\vspace{-0.08cm}}
\begin{equation}
(\!\Delta t_\pi\!)_{_{\rm B}} = (\:\!\!t_\pi\!)_{_{\rm B}} \!-\! t_\pi =
 -0.22 \; {\rm day}.
\end{equation}
In this case the observed comet is predicted to have reached perihelion
0.22~day earlier than in the absence of early fragmentation.  The result
may appear to be independent of the separation time, but in reality its
effect is hidden in the adopted values of $(\!\Delta t_\pi^\ast\!)_{\rm x}$,
as is obvious from both Table~3 and footnote~5.

Inserting the value of $(\!\Delta t_\pi\!)_{_{\rm B}}$ from Equation~(6)
into the appropriate line of Table~2 yields \mbox{$\Delta P_{_{\rm B}} =
+0.12$ year}, so that if the comet did {\it not\/} fragment far from~the~Sun,
the orbital period would have been --- on~the~\mbox{proposed} assumptions
--- slightly, by a little more than 6~weeks, {\it shorter\/}.  The
differential effect of the nongravitational acceleration on the original
and fragmented comet is calculated to be trivial.
%
% Our considerations have so far been confined to three companions.  We do
% not know the actual number of companions involved, nor do we know how
% far they stretched or when the parent of these sungrazers had begun
% to~frag\-ment.  A potential range of perihelion times of fragments is
% astonishing:\ a separation velocity of 1~m/s would cause a companion to
% arrive 5~months after the comet if detached at aphelion, but 2.7~years
% after the comet if detached at 100~au from the Sun on the way {\it to\/}
% aphelion,~to offer two examples.  The perihelion time of the observed
% comet would be affected greatly and so would be --- even though to a much
% lesser degree --- its orbital period.

A distinct and extremely interesting possibility is that comet C/2026~A1
together with the three bright companions were in fact the ``tip of the
iceberg,'' isolated pieces of an object, referred to from now on as {\it
Midget\/}, which --- following its separation from a massive parent in
AD~363 --- fragmented extensively at some point far from the Sun.  In
this scenario comet C/2026~A1 was merely Midget's {\small \bf largest
fragment} and the bright companions were the next most sizable fragments.
A large number of ever tinier sungrazers associated with them should have
been arriving over an extended period of time, nearly all following the
comet itself because of the nongravitational effects on their motions.

If the mass of C/2026~A1 was confined to not more than about one half
of Midget's mass, its motion could have been measurably perturbed when
Midget fell apart, and it appears that the odds were fairly high that
{\it Fragment~I\/}, the ancestor of the Great Comet of 1106 and the Great
March Comet of 1843 (cf.\ Figure 3 of Sekanina 2025),~was likewise the
{\small \bf parent} of both Midget and C/2026~A1.

Following Midget's separation at perihelion in AD~363, its orbit was
practically that of {\it Fragment~I\/} in the angular elements and
perihelion distance.  The only major difference was in the orbital
period, as explained elsewhere (Sekanina 2026a).  When Midget fragmented,
the angular elements of C/2026~A1 would change essentially on account
of the out-of-plane component, $V_{\rm N}$, of the separation velocity;
its perihelion distance largely because of the transverse component,
$V_{\rm T}$; and its time of the next perihelion and orbital period
mainly because of the radial component, $V_{\rm R}$.  When the observed
motion of C/2026~A1 is integrated back to the 4th century, these changes
remain of course unaccounted for, but they appear as differences in the
individual elements in AD~363 between C/2026~A1 and {\it Fragment~I\/},
the grandparent of C/1843~D1 (cf.\ Table 4).

We averaged the orbital elements of 38 clones passing perihelion in
AD~363 (22 from Run~v2, 16 from Run~v5) to approximate a modified
orbit of C/2026~A1, listed in Table~7.  Its comparison with the
363 orbit of C/1843~D1 in Table~4 provides the following differences,
in the sense C/2026~A1 minus C/1843~D1:\ \mbox{$\Delta \omega =
-2^\circ\!$.5},~\mbox{$\Delta \Omega = -2^\circ\!$.2}, \mbox{$\Delta
i = -0^\circ\!$.2}, and \mbox{$\Delta q = +0.00027$ au = +0.058
$R_\odot$}.~Fur\-ther, from the temporal separation between~C/2026~A1
and~the brightest observed companions it follows~that \mbox{$|\Delta
t_\pi| \ll 4$ days}.  For Midget's {\it assumed\/} fragmentation~at
aphelion, 280~au from the Sun, these differences~give~for
minimum values of the separation-velocity~components of C/2026 A1:\
\mbox{$V_{\rm N} \!=\! +0.31 \!\pm\! 0.05$ m/s},~\mbox{$V_{\rm T}
\!=\! +0.25$ m/s}, and \mbox{$|V_{\rm R}| \!\ll\! 0.03$ m/s}.

Similarly, a set of orbital elements by Nakano (personal communication via
D.\,W.\,E.\,Green), generated~by integrating the motion of C/2026~A1 and
forcing the perihelion time in November 363 (see Table~4;
also Sekanina 2026a), offers the following differences from the elements
of C/1843~D1 in Table~4:\ \mbox{$\Delta \omega = -3^\circ\!$.0},
\mbox{$\Delta \Omega = -3^\circ\!$.5}, \mbox{$\Delta i = -0^\circ\!$.3},
and \mbox{$\Delta q = +0.000353$ au = +0.076 $R_\odot$}.  The normal and
transverse components of the separation velocity for the breakup at
aphelion are now, respectively, $+0.44 \!\pm\! 0.04$~m/s and +0.33~m/s.

The assumption of Midget's fragmentation event taking place at aphelion
offers a useful constraint but is rather unlikely.  Its occurrence after
aphelion, 100--200~au from the Sun, with a somewhat higher separation
velocity, near 1~m/s, is more probable.  The orbital period~has in any
case been affected only marginally, having possibly changed by a fraction
of a year.

In summary, the very existence of the distant compan\-ions can in this
comprehensive scenario be construed as evidence of Midget's rather poor
cohesion, an inference that is broadly consistent with the failure of
C/2026~A1 to survive perihelion.  This actually was a warning \ldots

\section{Final Comments and Conclusions} % Section 8
The fiasco notwithstanding, comet C/2026 A1 was not a disappointment,
as it has greatly contributed to our understanding of a chain of
transformations that the Kreutz sungrazers could undergo during their
passage through perihelion.  In this sense, they could be divided into
five distinct categories.

Only a handful of objects are known to make up the first category.
Each was observed after perihelion as a single nebulosity that stayed
single.  These sungrazers include the Great March Comet\vspace{-0.035cm}
of 1843 (C/1843~D1), C/1880~C1, Pereyra (C/1963~R1),\footnote{Roemer
(1963, 1965) reported a possible secondary nucleus~0$^\prime\!$.1
from the primary on 1963 November~9, nearly 80~days after perihelion,
which was never confirmed.  Its existence is highly unlikely because
(i)~a genuine companion should have been detected much earlier, and
(ii)~this long after perihelion the separation distance from
the primary should have been substantially greater.} and\vspace{-0.035cm}
White-Ortiz-Bolelli (C/1970~K1).\footnote{We list only the objects
observed telescopically since the mid-19th century, because for
historical sungrazers the appearance of their nuclear condensation
is unknown.}  Although the 1843 comet was the most spectacular, it
too was fading rapidly after perihelion and, except for Pereyra,
all these objects were then observed over only fairly short periods
of time (the 1843 comet up to 51~days, the 1880 comet up to 23~days,
and the 1970 comet up to 24~days after perihelion).  Besides, Gould
(1891) reported that the observations of the 1880 comet ``were
rendered difficult by the lack of a nucleus or condensation in the
head, which appeared like a cloud, elongated in the direction of the
tail and of but slightly greater brilliancy.''  Similarly, on the last
day of its observation the 1970~comet was described by both M.\ V.\
Jones and A.\ F.\ Jones as faint, with no distinct head (Marsden 1971).
These comments suggest that the structural integrity of the nucleus may
have been compromised but the extent of damage could not be determined
because of inadequate spatial resolution.

The second category of Kreutz sungrazers includes brilliant objects
whose nuclei split at or close to perihelion into two or more
discrete major fragments and a lot of debris.  Two known
cases are the Great September Comet of 1882 (C/1882~R1), which
displayed up to six nuclear condensations after perihelion (e.g.,
Kreutz 1888, 1891),\footnote{The condensations were distributed in
a row, prompting Gill (1883) to remark that they resembled a line of
{\it small beads strung on a thread of worsted\/}.\vspace{-0.1cm}}
and Ikeya-Seki (C/1965~S1), which exhibited two persisting nuclei
(e.g., Marsden 1967) as well as possible additional fragments of
temporary nature (e.g., Hirayama \& Moriyama 1965).  Damage to the
nucleus is thus obvious.

The third category consists of less massive objects~that
survive$\:$perihelion$\:$but$\:$disintegrate$\:$shortly\,(hours$\:$to$\:$days)
afterwards, thereby generating an impressive dust tail.  The first
sungrazer of this kind was C/1887~B1; because of its appearance as
a tail with no head, it was commonly referred to as a ``headless
wonder.''  A study of records of the tail's path among the stars
over a period of 10~days suggested that the comet's entire mass
was smashed into micron- and submicron-sized dust by about 6~hours
after perihelion (Sekanina 1984).  Fifteen years ago the same kind
of event was displayed by another Kreutz comet, Lovejoy (C/2011~W3),
except that this time the nucleus disintegrated as late as
$\sim$40~hours after perihelion (Sekanina \& Chodas 2012).

Comet C/2026 A1 is a member of the fourth category of sungrazers:\
these are objects that fail to survive perihelion by a very
narrow margin.  The result is that the tail, made up of dust
ejected before perihelion, survives past the point of closest
approach to the Sun, but only very briefly.  It so happens
that submicron-sized dust particles, each moving in its own
hyperbolic orbit, avoid sublimation because their perihelion
distances are greater than the perihelion distance of the
nucleus.  Ironically, larger particles move in orbits whose
perihelion distances are nearer that of the nucleus and are
subjected to sublimation.  It is therefore only the outer
part of the comet's preperihelion tail that survives.  But
because the particles' hyperbolic orbits rapidly diverge, the
remnants of the tail fade promptly, over a period of a day
or two after perihelion.  The tail's projected shape depends
critically on the geometry.  In the case of C/2026~A1 the
Earth was close to the comet's orbital plane, so that the
cloud of debris acquired a geyser-like appearance.  Its
preliminary modeling showed that the formation of the tail
terminated about 2~hours before perihelion and that the peak
solar radiation-pressure acceleration, which the dust in the
cloud was subjected to, equaled 0.67 the solar gravitational
acceleration, implying the presence of silicate grains
(Sekanina 2026b).

Other Kreutz sungrazers whose preperihelion tail survived
both the head and perihelion point were C/1979~Q1 (Solwind~1,
also known as Howard-Koomen-Michels), the first ever comet
discovered with a space-borne instrument (Michels et al.\
1982); a bright dwarf SOHO sungrazer C/2007~L3 (Thompson
2009); and potentially additional bright dwarf sungrazers,
for which in-depth searches for a surviving pre\-perihelion
tail have never been conducted.  As a matter of curiosity
we point out that the images of C/2026~A1 in a STEREO-A
coronagraph, taken hours before perihelion, displayed
remarkable similarity to a photographic image of X/1882~K1
(Tewfik), exposed during the total solar eclipse on
1882~May~17 (Abney \& Schuster 1884).~We suggest that
this sungrazer might~also have belonged to this category.

Some of the disintegrating sungrazers with the preperihelion
tail surviving past perihelion are likely to have been
objects that were on collision course with the Sun.  A
primary suspect is C/1979~Q1, for which an orbit with
\mbox{$q > R_\odot$} was extremely hard to establish.
Eventually, Marsden (1989) barely succeeded, but only after
arbitrarily rotating the comet's Solwind-observed coordinates
by 12$^\circ$!  More generally, computations suggest that
a separation velocity's transverse component~of~1~m/s~near
aphelion would change the perihelion distance of a fragment
by 0.2$\,R_\odot$, so that collisions of dwarf sungrazers~with
the Sun certainly cannot be ruled out, especially
not given that multiple fragmentation events are implied
by their cascading nature.  Alternatively, Marsden (1989)
noted that under certain circumstances some sungrazers
could collide with the Sun because of the action of the
indirect planetary perturbations.

The fifth and last category includes an overwhelming
majority of the dwarf (largely SOHO) sungrazers, which
completely sublimate away before perihelion, mostly at
a time when they still have 0.2--0.3~day to go.  They
leave no surviving tails that could be detected in the
space-borne coronagraphic images.

The five categories of the Kreutz sungrazers correlate
with the degree of spectacle they display.  Noticeable
is a long-term trend, as the two most remarkable objects
appeared in the 19th century, while less impressive but
still brilliant ones were seen in the 20th century.  In
the 21st century we have so far been deprived of any
object of the first or second category, possibly an
indication of the fleeting nature of the Kreutz system.

Besides its taxonomic significance, C/2026 A1 stands out because
of its unique, unusually long orbital period.  Our baseline
hypothesis (Sekanina 2026a) maintains that the comet was a
small, outlying fragment of one of the brilliant comets
whose appearance in broad daylight in late AD~363 was
recorded by Ammianus Marcellinus, a noted Roman historian.
Accordingly, the object appears to be the only second-generation
fragment of Aristotle's comet known to have arrived after the
12th century.

In support of this hypothesis, we contemplated an extensive
investigation of the comet's orbital motion.  We first confirmed
that the 22~pre-discovery astrometric observations, starting on
2025 December~18 and reported by Deen (2026), were fully compatible
with the approximately 1000~post-discovery observations and helpful
in that they extended the observed orbital arc from 74~days to
100~days.  We suggested that the comet's motion appeared to have
measurably been affected by an abrupt change in the pattern of
outgassing, seen as an anomalous feature on the light curve.  The
increasing rapid brightening suddenly turned into a near-plateau,
perhaps a sign of the onset of gradual deactivation of a discrete
source on the nucleus.  The event occurred on February~\mbox{9--12},
less than 8~weeks before perihelion at a heliocentric distance of
1.55~au.  This apparently was the first in a sequence of episodes
that over a \mbox{6--7}~week interval influenced the comet's
motion to an extent that the computed orbital period grew as
much as 80~years, becoming more sensitive to the quirks of
activity than traditional orbital-noise measures.  The
observations over the entire orbital arc (2025 December~18--2026
March~28) could seemingly be satisfied by a gravitational solution,
even though sophisticated tests of the distribution of positional
residuals revealed systematic trends incompatible with the Gaussian
law and the amplitude of variations in the orbital period reached
levels that were one order of magnitude higher than the relevant
standard deviation (1$\sigma$).  For some unknown reason,
we were unable to reproduce the orbital periods computed,
respectively, by MPC and JPL from the observations covering
an interval of 2026 January~13 through March~28, as seen by
comparing Run~c6 from Table~5 with the last two entries of
Table~1.

Even though the problems with orbit determination appear
to have been triggered by outgassing-driven nongravitationa
forces, nongravitational solutions did not do better than
the gravitational solutions.  We believe that the culprit
was the erratic nature of the forces, so that the applied
nongravitational function, be it the standard law $g(r)$
or a similar one, was incapable of successfully approximating
the genuine variations.

We derived two gravitational and two nongravitational solutions
based on observations terminated by the February anomalous
feature on the light curve.  Two options were necessitated
by uncertainties (a)~in this anomaly's timing; and (b)~stemming
from~data~noise~along~the~pre\-cariously short length of the
truncated arc, which extended over less than two months.
The end date for one option, Runs~v5 (gravitational) and v6
(nongravitational), was February~9; for the other option, Runs~v2
(gravitational) and v3 (nongravitational), it was February~12.  The
beginning was the same, 2025 December~18, the date of the first
pre-discovery observation.

Although uncertain to about $\pm$15~years, the dates of the previous
perihelion derived from the gravitational solutions --- Runs~v5 and
v2 --- came out within $\sim \!\! 1 \sigma$ of the time of appearance
of the daylight comets recorded by Ammianus Marcellinus.  We judge
these sets of elements (Table~7) the best approximations to the
genuine orbit of comet C/2026~A1 before it was being continuously
modified by appreciable outgassing-driven nongravitational forces
of erratic nature.  The two nongravitational solutions failed to
offer meaningful results apparently because the orbital arc was
much too short.

The existence of distant companions, a problem that needs more
future attention, suggests that C/2026~A1 separated from one of
the spectacular comets at perihelion in AD~363 only as part of
another object --- called here Midget --- that much later fell
apart, C/2026~A1 being its largest fragment.  Even though we
have~no~di\-rect evidence of the event itself, the relationship
between the orbits of C/2026~A1 and {\it Fragment~I\/}, the
ancestor of C/1843~D1 in AD~363, substantiating this scenario,
is supported by both our computations and those performed by
Nakano.  If {\it Fragment~I\/} was the spectacular comet that
Midget separated from, their elements --- except for the orbital
period --- must have been the same when leaving the 363 perihelion.
The impact of Midget's fragmentation on the motion of C/2026~A1 long
after its separation is seen from the differences between the~orbits
of C/2026~A1 and C/1843~D1, both integrated back to AD~363.  On an
extreme assumption that Midget fragmented at aphelion, 280~au from
the Sun, the differences in the tabulated angular elements and
perihelion distance imply a submeter-per-second velocity to separate
C/2026~A1 from Midget.~In~reality,~it~was~more~likely that Midget
fragmented way after aphelion and the separation velocity was near
1~m/s.  In any case, as a piece~of {\it Fragment~I\/}, comet
C/2026~A1 was member of Population~I, associated with the Great March
Comet of 1843.

% In summary, \ldots\ldots (final statement on the nature and identity
% of the comet). {\bf To be continued}. \ldots\ldots
%
\vspace*{0.2cm}
\begin{center}
{\footnotesize REFERENCES}
\end{center}
\vspace*{0.1cm}
% \hspace{-0.39cm}
%
\parbox{8.63cm}{\footnotesize
Abney, W.\ de W., \& Schuster, A.\ 1884, Phil.\ Trans.\ Roy.\ Soc.
 {\hspace*{0.25cm}}London, 175, 253 \\[0.03cm]
%
% Bailey,\,M.\,E.,\,Chambers,\,J.\,E.,\,\&\,Hahn,\,G.\,1992,\,Astron.\,Astrophys.,
% {\hspace*{0.25cm}}257, 315 \\[0.03cm]
%
Bielicki, M., \& Sitarski, G.\ 1991, Acta Astron., 41, 309 \\[0.03cm]
Deen, S.\ 2026, Centr.\ Bur.\ Electr.\ Tel.\ No.\ 5658 \\[0.03cm]
Dybczy{\'n}ski, P.\ A., \& Kr\'olikowska, M.\ 2025, Astron.\ Astrophys.,
 {\hspace*{0.25cm}}702, A143 \\[0.03cm]
%
% England, K.\ J.\ 2002, J.\ Brit.\ Astron.\ Assoc., 112, 13 \\[0.03cm]
%
% Fabritius, W.\ 1883, Astron.\ Nachr., 105, 287 \\[0.03cm]
%
% Frisby, E.\ 1883, Astron.\ Nachr., 104, 159 (erratum:\ 283); also:\ Na-
%  {\hspace*{0.25cm}}ture, 27, 226 \\[0.03cm]
%
Gill, D.\ 1883, Mon.\ Not.\ Roy.\ Astron.\ Soc., 43, 319 \\[0.03cm]
Gould, B.\ A.\ 1891, Result.\ Obs.\ Nacion.\ Arg., 13, 600 \\[0.03cm]
Green, D.\ W.\ E.\ 2026a, Centr.\ Bur.\ Electr.\ Tel.\ No.\ 5658 \\[0.03cm]
Green, D.\ W.\ E.\ 2026b, Centr.\ Bur.\ Electr.\ Tel.\ No.\ 5675 \\[0.03cm]
%
% Hasegawa, I.\ 1980, Vistas Astron., 24, 59 \\[0.03cm]
%
% Hasegawa, I., \& Nakano, S.\ 2001, Publ.\ Astron.\ Soc.\
% Japan,~53,~931 \\[0.03cm]
%
Hirayama, T., \& Moriyama, F.\ 1965, Publ.\ Astron.\ Soc.\ Japan, 17,
 {\hspace*{0.25cm}}433 \\[0.03cm]
%
% Ho, P.-Y.\ 1962, Vistas Astron., 5, 127 \\[0.03cm]
%
% Holetschek, J.\ 1892, Astron.\ Nachr., 129, 323 \\[0.03cm]
%
% Hubbard, J.\ S.\ 1849, Astron.\ J., 1, 10 \\[0.03cm]
%
% Hubbard, J.\ S.\ 1850, Astron.\ J., 1, 153 \\[0.03cm]
%
% Hubbard, J.\ S.\ 1851a, Astron.\ J., 2, 46 \\[0.03cm] % cape obs received
%
% Hubbard, J.\ S.\ 1851b, Astron.\ J., 2, 57 \\[0.03cm] % orbit VI
%
% Hubbard, J.\ S.\ 1852, Astron.\ J., 2, 153 \\[0.03cm] % orbit VII
%
% Hufnagel, L.\ 1919, Astron.\ Nachr., 209, 17 \\[0.03cm]
%
Knight,\,M.\,M., A'Hearn,\,M.\,F., Biesecker,\,D.\,A., et al.\,2010,~\mbox{Astron}.
 {\hspace*{0.25cm}}J., 139, 926 \\[0.03cm]
Kreutz, H.\ 1888, Publ.\ Sternw.\ Kiel, No.\ 3 \\[0.03cm]
Kreutz, H.\ 1891, Publ.\ Sternw.\ Kiel, No.\ 6 \\[0.03cm]
%
% Kreutz, H.\ 1895, Astron.\ Nachr., 139, 113 \\[0.03cm]
%
Kreutz, H.\ 1901, Astron.\ Abhandl., 1, 1 \\[0.03cm]
Kr\'olikowska, M., \& Dones, L.\ 2023, Astron.\ Astrophys., 678, A113 \\[0.03cm]
Kr\'olikowska, M., \& Dybczy{\'n}ski, P.\,A.\ 2010, Mon.\,Not.\,Roy.\,Astron.
 {\hspace*{0.25cm}}Soc., 404, 1886 \\[0.03cm]
Kr\'olikowska, M., Sitarski, G., \& So{\l}tan, A.\,M.\ 2009, Mon.\,Not.\,Roy.
 {\hspace*{0.25cm}}Astron.\ Soc., 399, 1964 \\[0.03cm]
%
% Landgraf, W.\ 1985, Sterne, 61, 351 \\[0.03cm]
%
% Maclear, T.\ 1851, Mem.\ Roy .\ Astron.\ Soc., 20, 62 \\[0.03cm]
%
Marcus, J.\ N.\ 2007, Int.\ Comet Quart., 29, 39 \\[0.03cm]
Marsden, B.\ G.\ 1967, Astron.\ J., 72, 1170 \\[0.03cm]
Marsden, B.\ G.\ 1971, Quart.\ J.\ Roy.\ Astron.\ Soc., 12, 244 \\[0.03cm]
Marsden, B.\ G.\ 1989, Astron.\ J., 98, 2306 \\[0.03cm]
Marsden, B.\ G., \& Sekanina, Z.\ 1978, Astron.\ J., 83, 64 \\[0.03cm]
%
% Marsden, B.\ G., \& Williams, G.\ V.\ 2008, Catalogue of Cometary
% {\hspace*{0.25cm}}Orbits 2008 (17th ed.).  Cambridge, MA:\ IAU Central
% Bureau~for
% {\hspace*{0.25cm}}Astronomical Telegrams and Minor Planet Center,
% 195pp\\[0.03cm]
%
% Marsden,\,B.\,G., Sekanina,\,Z., \& Everhart,\,E. 1978,
% Astron.\,J.,~83,~64 \\[0.03cm]
%
Marsden, B.\ G., Sekanina, Z., \& Yeomans, D.\ K.\ 1973, Astron.~J.,
 {\hspace*{0.25cm}}78,~211 \\[0.03cm]
%
% Mart\'{\i}nez, M.\ J., Marco, F.\ J., Sicoli, P., \& Gorelli, R.\ 2022,~Icarus,
% {\hspace*{0.25cm}}384, 115112 \\[0.03cm]
%
Maury, A.\ 2026, Centr.\ Bur.\ Electr.\ Tel.\ No.\ 5658 \\[0.03cm]
%
% Meyer, M., \& Kronk, G.\ W.\ 2026, J.\ Astron.\ Hist.\ Herit., 29, 93 \\[0.03cm]
%
Michels, D.\ J., Sheeley, Jr., N.\ R., Howard, R.\ A., \& Koomen, M.~J.\
 {\hspace*{0.25cm}}1982, Science, 215, 1097 \\[0.03cm]
%
% Morrison, J.\ 1883, Mon.\ Not.\ Roy.\ Astron.\ Soc., 44, 49 \\[0.03cm]
%
Nakano, S.\ 2026a, Centr.\ Bur.\ Electr.\ Tel.\ No.\ 5658 \\[0.03cm]
Nakano, S.\ 2026b, NK 5553 \\[0.03cm]
Nakano, S.\ 2026c, NK 5561 \\[0.03cm]
Nakano, S.\ 2026d, NK 5566 \\[0.03cm]
%
% Pingr\'e,\ A.\ G.\ 1783, Com\'etographie ou Trait\'e Historique et Th\'eo-
% {\hspace*{0.25cm}}rique des Com\`etes. Paris:\ L'Imprimerie Royale,
% 630pp \\[0.03cm]
%
Roemer, E.\ 1963, Publ.\ Astron.\ Soc.\ Pacific, 75, 535 \\[0.03cm]
Roemer, E.\ 1965, Astron.\ J., 70, 397 \\[0.03cm]
Sekanina, Z.\ 1984, Icarus, 58, 81 \\[0.03cm]
Sekanina, Z.\ 2000, Astrophys.\ J., 542, L147 \\[0.03cm]
Sekanina, Z.\ 2021, eprint arXiv:2109.01297 \\[0.03cm]
Sekanina, Z.\ 2025, eprint arXiv:2503.15467 \\[0.03cm]
Sekanina, Z.\ 2026a, eprint arXiv:2602.17626 \\[0.03cm]
Sekanina, Z.\ 2026b, Centr.\ Bur.\ Electr.\ Tel.\ No.\ 5681 \\[0.03cm]
Sekanina, Z., \& Chodas, P.\ W.\ 2012, Astrophys.\ J., 757, 127 (33pp) \\[0.03cm]
Sekanina, Z., \& Kracht, R.\ 2013, Astrophys.\ J., 778, 24 (13pp) \\[0.03cm]
Sekanina, Z., \& Kracht, R.\ 2022, eprint arXiv:2206.10827 \\[0.03cm]
Sitarski, G.\ 1998, Acta Astron., 48, 547 \\[0.03cm]
Thompson, W.\ T.\ 2009, Icarus, 200, 351 \\[0.03cm]
Zhang, Q., Knight, M.\ M., Ye, Q., et al.\ 2026, Res.\ Notes Amer.\
 {\hspace*{0.25cm}}Astron.\ Soc., 10, 57}
\vspace{-0.08cm}
\end{document}